\documentclass[12pt,preprint]{aastex}
\def\degree{\hbox{$^\circ$}}

\usepackage{amssymb,amsmath}
\usepackage[usenames,dvipsnames]{color}          % For color text: \color command

\shorttitle{Conversion to Alfv\'en waves in sunspots}

\shortauthors{Khomenko \& Cally}

\begin{document}

\title{Numerical simulations of conversion to Alfv\'en waves in sunspots}

\author{E.~Khomenko\altaffilmark{1,2} and P.~S.~Cally\altaffilmark{3}}
\email{khomenko@iac.es}
\email{paul.cally@monash.edu}
\altaffiltext{1}{Instituto de Astrof\'{\i}sica de Canarias, 38205,
C/ V\'{\i}a L{\'a}ctea, s/n, La Laguna, Tenerife, Spain}
\altaffiltext{2}{Departamento de Astrof\'{\i}sica, Universidad de
La Laguna, 38205, La Laguna, Tenerife, Spain}
\altaffiltext{3}{School of Mathematical Sciences and Monash Centre
for Astrophysics, Monash University , Clayton, Victoria, 3800 ,
Australia }

\begin{abstract}
We study the conversion of fast magneto-acoustic waves to Alfv\'en
waves by means of 2.5D numerical simulations in a sunspot-like
magnetic configuration. A fast, essentially acoustic, wave of a
given frequency and wave number is generated below the surface and
propagates upward though the Alfv\'en/acoustic equipartition layer
where it splits into upgoing slow (acoustic) and fast (magnetic)
waves. The fast wave quickly reflects off the steep Alfv\'en speed
gradient, but around and above this reflection height it partially
converts to Alfv\'en waves, depending on the local relative
inclinations of the background magnetic field and the wavevector.
To measure the efficiency of this conversion to Alfv\'en waves we
calculate acoustic and magnetic energy fluxes. The particular
amplitude and phase relations between the magnetic field and
velocity oscillations help us to demonstrate that the waves
produced are indeed Alfv\'en waves. We find that the conversion to
Alfv\'en waves is particularly important for strongly inclined
fields like those existing in sunspot penumbrae. Equally important
is the magnetic field orientation with respect to the vertical
plane of wave propagation, which we refer to as ``field azimuth''.
For field azimuth less than 90{\degree} the generated Alfv\'en
waves continue upwards, but above 90{\degree} downgoing Alfv\'en
waves are preferentially produced. This yields negative Alfv\'en
energy flux for azimuths between 90\degree\ and 180\degree.
Alfv\'en energy fluxes may be comparable to or exceed acoustic
fluxes, depending upon geometry, though computational exigencies
limit their magnitude in our simulations.
\end{abstract}


\section{Introduction}

Observations using the Solar Optical Telescope (SOT) aboard
\emph{Hinode} \citep{De-McICar07aa} and the Coronal Multi-Channel
Polarimeter (CoMP) \citep{TomMcIKei07aa} unambiguously reveal
ubiquitous Alfv\'enic oscillations in the solar corona, with
implications both for the Sun's atmosphere and the solar wind.
\citeauthor{TomMcIKei07aa} find that coronal Alfv\'enic power is
broadly spread in frequency, but with a distinct peak around 3-4
mHz, characteristic of the Sun's internal $p$-mode wavefield.
Whether these transverse oscillations are strictly Alfv\'en waves
or instead kink waves \citep{VanNakVer08aa} depends on the
magnetic and density structuring of the atmosphere, and may vary
with height as cross-field structuring becomes more or less
important. We might expect though that sunspot atmospheres, with
their presumed large-scale
%relatively smooth
magnetic structures,
present the most ideal site for more-or-less pure Alfv\'en waves
to propagate.

The generation of Alfv\'en waves at the photosphere and their
propagation through the various layers of the solar atmosphere has
been extensively modelled \citep[e.g.][]{Cravan05aa}. However,
\cite{CallyHansen2011} recently suggested that Alfv\'en waves must
also be produced by mode conversion from fast magneto-acoustic
waves beyond their reflection height in the low chromosphere. This
coupling only occurs if wave propagation is not co-planar with
gravity and magnetic field, and so the problem is necessarily
three dimensional (3D).

Conversion from fast-mode high-$\beta$ magneto-acoustic waves
(manifesting as $p$ modes in the subphotosphere) to slow-mode
waves in solar active regions is relatively well studied both
analytically and numerically
\cite[e.g.,][]{Zhugzhda+Dzhalilov1982, Cally+Bogdan1997,
Cally2001, Schunker+Cally2006, Cally2006, Cally2007, HanCal09aa,
Khomenko+etal2009, Felipe+etal2010a}; see \citet{Khomenko2009} for
a review. In a two-dimensional situation, the transformation from
fast to slow magneto\-acoustic modes is demonstrated to be
particularly strong for a narrow range of magnetic field
inclinations around 20--30 degrees to the vertical. For this
reason, and because of the reduction in acoustic cutoff frequency
afforded by strong inclined magnetic fields, magnetic field
concentrations on the solar surface may truly be called
\emph{magnetoacoustic portals} \citep{Jefferies+etal2006},
coupling the Sun's interior oscillations to those of its
atmosphere.

The remainder of the wave-energy flux though, which is near-total
away from these preferred inclinations, enters the low-$\beta$
atmosphere as fast, predominantly magnetic waves. Due to the steep
Alfv\'en speed gradient with height, these fast waves soon reflect
back downward at the height $z_{\rm ref}$ at which $\omega\approx
v_A k_h$, where $\omega$ is the frequency, $v_A$ the Alfv\'en
speed, and $k_h$ the horizontal wavenumber.
%
%Above this the fast waves are evanescent.

\begin{figure*}
\center
\includegraphics[width=.9\hsize]{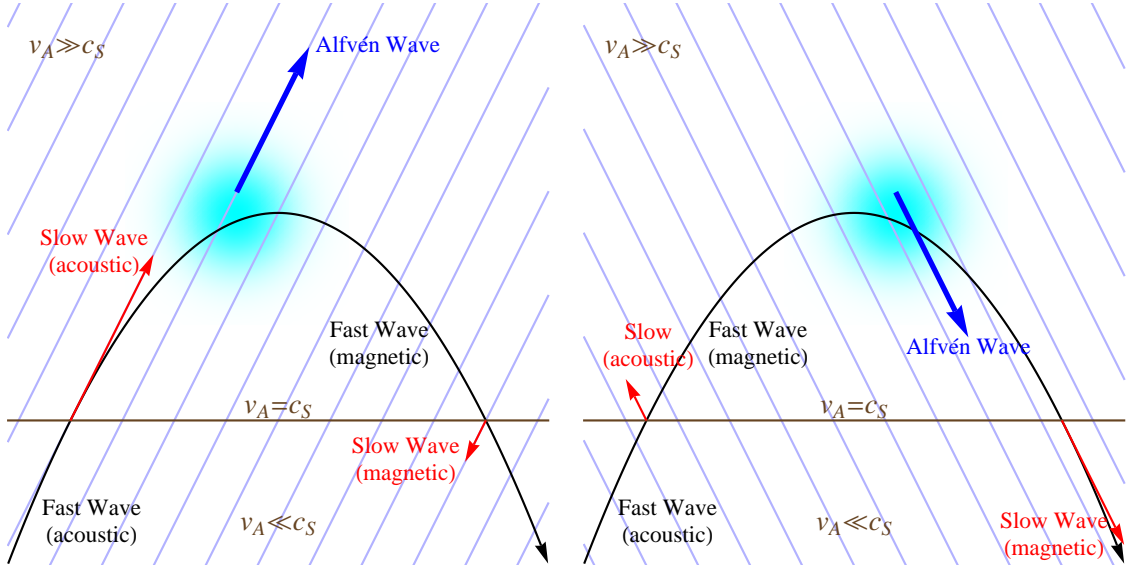}
\caption{Schematic diagram illustrating the various mode
conversions and reflections as a seismic ray (labeled ``Fast Wave
(acoustic)'') enters the solar atmosphere in a region of strong
inclined magnetic field. Field lines (pale blue) are oriented out
of the plane and are shown here in projection. Their orientation
is given by the inclination angle $\theta$ from vertical ($0 \le
\theta < 90^\circ$) and azimuth angle $\phi$ measured clockwise
from the wave propagation plane. By symmetry we need only consider
$0\le\phi\le180^\circ$, with $\phi<90^\circ$ in the left panel and
$\phi>90^\circ$ in the right panel. First, at the
Alfv\'en-acoustic equipartition level $v_A=c_S$ the ray splits
into an essentially acoustic field-guided slow wave (depicted in
red) and a fast magnetically dominated wave (black). The slow wave
may or may not reflect depending on whether
$\omega<\omega_c\cos\theta$. The fast wave goes on to reflect
higher in the atmosphere due to the rapidly increasing Alfv\'en
speed with height. On its way downward it again mode converts at
the equipartition level. In the scenario depicted on the left
($\phi < 90^\circ$), the upward slow wave is much stronger than
the downward one because the fast ray is more closely aligned with
the magnetic field (small attack angle) on the upstroke than on
the downstroke. This situation is reversed if the magnetic field
were inclined in the opposite direction (equivalent to $\phi >
90^\circ$, right panel). In a nebulous region around and above the
fast wave reflection point (which may extend far higher than the
fuzzy blob used to represent it here), fast-to-Alfv\'en conversion
occurs, in the case on the left predominantly to an upgoing
Alfv\'en wave. For the $\phi > 90^\circ$ (right), the downgoing
Alfv\'en wave is favored. The fast-to-Alfven conversion may only
occur where the wave vector and the magnetic field lines are not
in the same vertical plane. This diagram extends the description
in \citet{Cally2007} by including fast-to-Alfv\'en
conversion.}\label{fig:AlfSchem}
\end{figure*}

Fast-to-Alfv\'en conversion occurs around and above this fast wave
reflection height $z_{\rm ref}$ in 3D (see
Fig.~\ref{fig:AlfSchem}), localized more closely to $z_{\rm ref}$
as frequency increases \citep{CallyHansen2011}. However, at the
3-5 mHz frequencies characteristic of $p$-modes, the process is
typically spread over much of the chromosphere. Studies of
fast-to-Alfv\'en conversion were initiated by
\citet{Cally+Goossens2008}, who found that it is most efficient
for preferred field inclinations from vertical between 30 and 40
degrees, and azimuth angles (the angle between the vertical
magnetic and wave propagation planes) between 60 and 80 degrees,
and that Alfv\'enic fluxes transmitted to the upper atmosphere can
exceed acoustic fluxes in some cases. \citet{Newington+Cally2010}
studied the conversion properties of low-frequency ($\sim1$-2 mHz)
gravity waves, showing that even larger magnetic field
inclinations can support gravity wave to Alfv\'en conversion and
resulting Alfv\'enic wave propagation to the upper atmosphere.
In nonuniform magnetic field, the relevant angles for mode
conversion, either fast-to-slow or fast-to-Alfv\'en, are those
pertaining in the respective conversion regions.
Fig.~\ref{fig:AlfSchem} summarizes the overall picture of
conversion between the fast, slow and Alfv\'en waves.

Motivated by these recent studies, we attack the problem by
means of 2.5D numerical simulations. The purpose of our analysis is
to calculate the efficiency of the conversion from fast-mode
high-$\beta$ magneto-acoustic waves to Alfv\'en and slow waves in
the upper atmosphere in a spreading sunspot-like magnetic field
configuration.
Our initial results on the conversion to Alfv\'en waves in simple
field configurations were reported in \citet{Khomenko+Cally2011}.
In the present work we extend our simulations to a more realistic
case of a sunspot atmospheric models spanning the shallow
subphotosphere ($-5$ Mm) to the high chromosphere ($+1.9$ Mm).

Above this layer, the transition region acts as a partial Alfv\'en
wave reflector, affecting their distribution in the corona and
solar wind and producing Alfv\'en turbulence via nonlinear
coupling \citep{Cravan05aa}, but
%for numerical reasons
this is beyond the scope of our present modeling.

\begin{figure*}
\center
\includegraphics[width=18cm]{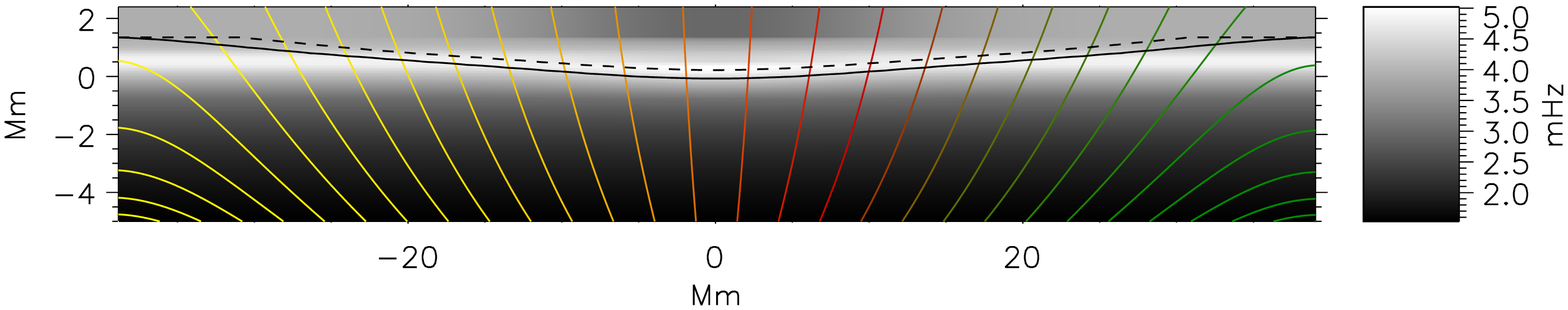}
\caption{Properties of the magneto-static sunspot model in the
$Y=7.5$ Mm slice away from the axis. The background image is the
acoustic cut-off frequency unadjusted for magnetic effects. The
solid black line is the level where acoustic and Alfv\'en speeds
are equal ($c_S=v_A$). The dashed line is the fast mode reflection
level calculated as the level where the wave frequency $\omega$
and wave horizontal wave number $k_x$ are related by
$\omega=v_Ak_x$. The magnetic field lines are shown as colored
lines, different colors meaning different azimuth $\phi$ of the
field from green ($\phi \approx 0^\circ$) to red ($\phi \approx
90^\circ$) and yellow ($\phi \approx 180^\circ$);  the $\phi$ is
counted clockwise from the vertical wave propagation plane XZ.}
Vertical and horizontal dimensions are not to scale.
\label{fig:sunspot}
\end{figure*}

%%%%%%%%%%%%%%%%%%%%%%%%%%%%%%%%%%%%
\section{Simulation setup}

We numerically solve the non-linear equations of ideal MHD using
our code \texttt{Mancha} \citep{Khomenko+Collados2006,
Khomenko+etal2008, Felipe+etal2010a}. The code solves non-linear
equations for perturbations, the equilibrium state being
explicitly removed from the equations.
In this study, for simplicity, and coherence with previous studies
\citep[see][]{CallyHansen2011}, we use a 2.5D approximation. This
approximation means that we allow all vectors in three spatial
directions, but the derivatives are taken only in two (one
vertical and one horizontal) directions, so that our perturbations
are only allowed to propagate in the $X-Z$ plane. In addition, the
initial perturbation is kept small to approximate the linear
regime.

As a background magneto-static model atmosphere we use one
sunspot-like model from \citet{Khomenko+Collados2008}. For
simplicity, this model is azimuthally symmetric, with no twist of
the magnetic field lines. Whilst twist is often seen in sunspot
magnetic fields, it is not a necessary feature of the processes we
wish to explore here. It will be added along with other physical
features in later studies. The computational region is magnetized
in all its volume (distributed currents), but the strongest
magnetic field is concentrated around the axis of the structure.
The dimensions of the simulated domain are 78 Mm in the horizontal
$X$ direction and 7.4 Mm in the vertical $Z$ direction. The axis
of the sunspot is placed at the middle of the domain  at $X=39$
Mm. The bottom boundary of the domain is located at $-5$ Mm below
the photospheric level, $Z=0$. This zero level is taken to be the
height where the optical depth at 500 nm is equal to unity in the
quiet Sun atmosphere 39 Mm away from the sunspot axis. The
thermodynamic variables of the atmosphere at 39 Mm from the axis
are taken from Model S of \citet{Christensen-Dalsgaard+etal1996}
in the deep sub-photosphere layers and continuing according to the
VAL-C model \citep{Vernazza+Avrett+Loeser1981} in the photospheric
and chromospheric layers. The sunspot axis in the atmospheric
layers is given by the semi-empirical model of \citet{Avrett1981}.
The magnetic field at the axis is about 900 G at $Z=0$ Mm. This is
quite moderate for a sunspot, but is adopted for numerical
reasons. The Alfv\'en speed at the top of our computational domain
becomes quite excessive if the field strength is too high, thereby
necessitating impractically small time steps. Taking larger field
strength would mostly result in a different scaling of the
problem, without new physical phenomena being introduced
\citep[see, e.g.][]{Khomenko+etal2009}.

The spatial resolution of our simulations is 150 km in the
horizontal direction and 50 km in the vertical direction (520 by
148 grid points). The upper 500 km (10 grid points) of the domain
are occupied by the absorbing PML boundary layer
\citep{Khomenko+etal2008, Felipe+etal2010a}, so the effective
physical boundary of the domain is located at 1.9 Mm above the
photosphere. The PML, ``Perfectly Matched Layer'', is a numerical
device that acts as an excellent absorber of waves, and thereby
removes unphysical reflections from the edges of the computational
domain. It is well-tested in the current code, and performs
admirably. No significant reflections are detected. The concept of
PML was firstly introduced by \citet{Berenger1994} for
electromagnetic waves, but was quickly extended to other wave
types. It is now used in many codes modeling wave propagation,
e.g. \citet{Parchevsky+Kosovichev2007} and \citet{Hanasoge2010}.
The calculations throughout are adiabatic and we use the ideal
equation of state to close the system.

As our modelling is 2.5D, we have made cuts through the sunspot
model at several $Y$ positions. Here we will describe three
simulations situated in vertical planes at $Y=7.5$, $11.25$ and
$15$ Mm from the sunspot axis. A cut through the axis is
uninteresting, as it is strictly 2D and can produce no Alfv\'en
waves. Figure \ref{fig:sunspot} depicts variations of some
characteristic parameters of the $Y=7.5$ Mm cut through the
sunspot model.

We drive waves by an imposed perturbation in a few grid points
near the bottom boundary of the domain at $Z=-5$ Mm. The
perturbation is calculated analytically as an acoustic-gravity
wave of a given frequency and wavenumber, neglecting the magnetic
field (dynamically unimportant at this depth), and neglecting the
temperature gradient. We introduce self-consistent perturbations
of the velocity vector, pressure, and density according to
\cite{Mihalas+Mihalas1984}:
\begin{equation}
\label{eq:initial}
\begin{split}
\delta V_{z} &  =  V_0 \exp\left( \frac{z}{2H}+k_{zi}z
\right)\sin(\omega t - k_{zr} z - k_x x) \\ %\nonumber
\frac{\delta P}{P_0} & =  V_0|P| \exp\left( \frac{z}{2H}+k_{zi}z
\right)\sin(\omega t - k_{zr} z -k_x x + \phi_P) \\ %\nonumber
\frac{\delta\rho}{\rho_0} & = V_0 |R| \exp\left(
\frac{z}{2H}+k_{zi}z \right)\sin(\omega t - k_{zr} z -k_x x +
\phi_R) \\ % \nonumber
\delta V_{x} & = V_0 |U| \exp\left( \frac{z}{2H}+k_{zi}z
\right)\sin(\omega t - k_{zr} z - k_x x + \phi_U)\,,
\end{split}
\end{equation}
where the amplitudes and relative phase shifts between the
perturbations are given by
\begin{gather}
|P|=\frac{\gamma\omega}{\omega^2 - c_S^2k_x^2}\sqrt{k_{zr}^2 +
\left( k_{zi} + \frac{1}{2H}\frac{(\gamma - 2)}{\gamma} \right)^2}\\
|R|=\frac{\omega}{\omega^2 - c_S^2k_x^2}\sqrt{k_{zr}^2 + \left(
k_{zi} + \frac{(\gamma - 1)}{\gamma H}\frac{c_S^2k_x^2}{\omega^2}
- \frac{1}{2H} \right)^2}\\
|U|=\frac{k_xc_S^2}{\gamma\omega}|P|\\
\phi_P = \phi_U = \arctan\left( \frac{k_{zi}}{k_{zr}} +
\frac{1}{2Hk_{zr}}\frac{(\gamma - 2)}{\gamma}   \right)\\
\phi_R =\arctan\left(  \frac{k_{zi}}{k_{zr}} + \frac{(\gamma -
1)}{\gamma Hk_{zr}}\frac{c_S^2k_x^2}{\omega^2} -
\frac{1}{2Hk_{zr}} \right)
\end{gather}

Given the wave frequency $\omega$, and the horizontal wavenumber
$k_x$, the vertical wavenumber $k_z$ is found from the dispersion
relation for acoustic-gravity waves in an isothermal atmosphere:
\begin{equation}
k_z=k_{zr} + ik_{zi} = \sqrt{(\omega^2-\omega_c^2)/c_S^2 -
k_x^2(\omega^2-\omega_g^2)/\omega^2}
\end{equation}
where $\omega_c=\gamma g/2 c_S$ is the acoustic cut-off frequency and
$\omega_g=2\omega_c\sqrt{\gamma-1}/\gamma$.

In all simulations described in this work we set the perturbation
frequency $\nu=\omega/2\pi=5$ mHz and horizontal wave number
$k_x=1.37$ Mm$^{-1}$. According to our sunspot model (see
Fig.~\ref{fig:sunspot}), the driving frequency is just slightly
below the maximum cut-off frequency reached at the temperature
minimum. Over most of the spot these values result in the fast
wave reflection level $z_{\rm ref}$ being higher than the
Alfv\'en/acoustic equipartition surface $z_{\rm eq}$, i.e.,
$\omega/k_x>c_S$ (automatically satisfied for a propagating
acoustic wave $\omega^2=c_S^2(k_x^2+k_z^2)$). This is important,
since it allows upcoming acoustic (fast) waves in $v_A<c_S$ to
convert to magnetic (fast) waves in $v_A>c_S$ before reflecting at
$z_{\rm ref}$.

%It is useful to note that in an isothermal atmosphere with
%density scale height $H$, $z_{\rm ref}-z_{\rm eq}=2H\ln(\omega/c_Sk_x)$,
%and so $z_{\rm ref}>z_{\rm eq}$ provided that $\omega>c_Sk_x$, which is
%of course satisfied for a vertically propagating acoustic wave,
%for which $\omega^2=c_S^2(k_x^2+k_z^2)$.

To separate the Alfv\'en mode from the fast and slow
magneto-acoustic modes in the magnetically dominated atmosphere
($v_A \gg c_S$) we use velocity projections onto three
characteristic directions:
\begin{equation}
\label{eq:directions}
\begin{split}
 \hat{e}_{\rm long} & =  [\cos\phi
\sin\theta, \, \sin\phi \sin\theta, \, \cos\theta];   \\
\hat{e}_{\rm perp} & =  [  -  \cos\phi \sin^2\theta \sin\phi, \,
1-\sin^2\theta \sin^2\phi,  \\ &\quad -   \cos\theta
\sin\theta \sin\phi]; \\
\hat{e}_{\rm trans} & =  [-\cos\theta, \, 0, \, \cos\phi
\sin\theta].
\end{split}
\end{equation}
These projections were shown to be rather efficient in separating
the perturbations corresponding to all three modes both for
idealized magnetic field configurations
\citep{Khomenko+Cally2011}, and also for more complex ones
\citep{Felipe+etal2010a}. The first projection ($\hat{e}_{\rm
long}$) selects the slow magneto-acoustic wave, propagating parallel
to the field; the second one ($\hat{e}_{\rm perp}$) selects the
Alfv\'en wave, according to the asymptotic polarization direction
derived by \citet{Cally+Goossens2008}; the remaining orthogonal
direction ($\hat{e}_{\rm trans}$) selects the fast magneto-acoustic
wave. In the rest of the paper we will address these
projections as ``acoustic'', ``Alfv\'en'' and ``fast''.

\begin{figure*}
\center
\includegraphics[width=18cm]{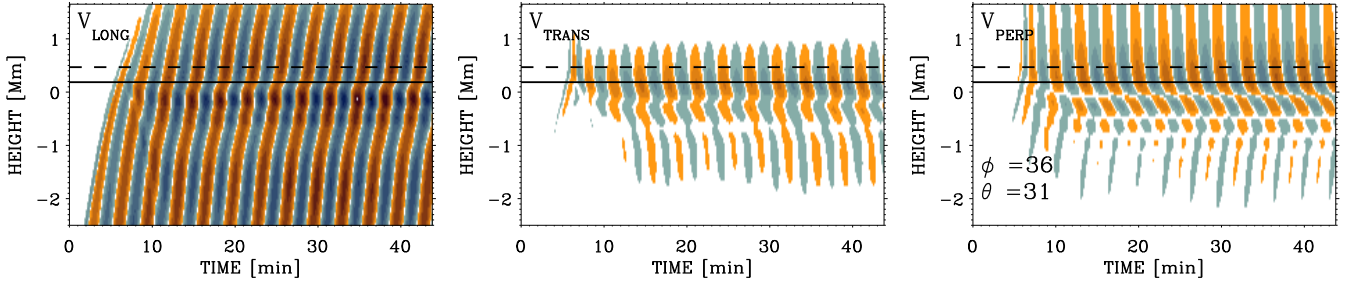}
\caption{Variations with time (horizontal axis) and height
(vertical axis) of the three orthogonal components of the velocity
in the simulation at $Y=7.5$ Mm from the sunspot axis. The color
coding is the same in the three panels. The velocities are scaled
with a factor of $\sqrt{\rho_0 c_S}$ (left panel) and
$\sqrt{\rho_0 v_A}$ (middle and right panels). Horizontal solid
line is the level where $c_S=v_A$; horizontal dashed line is the
fast mode reflection level. Velocities are taken at one horizontal
$X$ location where the magnetic field inclination
$\theta=31$\degree\ and azimuth $\phi=36$\degree. While the fast
mode (visible in $V_{\rm trans}$ above the $c_S=v_A$ level) is
reflected, the Alfv\'en mode (visible in $V_{\rm perp}$ above the
$c_S=v_A$ level) continues above. Note the difference in the
propagation speeds of the acoustic and Alfv\'en wave above
$c_S=v_A$. }\label{fig:vtime}
\end{figure*}

\begin{figure*}
\center
\includegraphics[width=18cm]{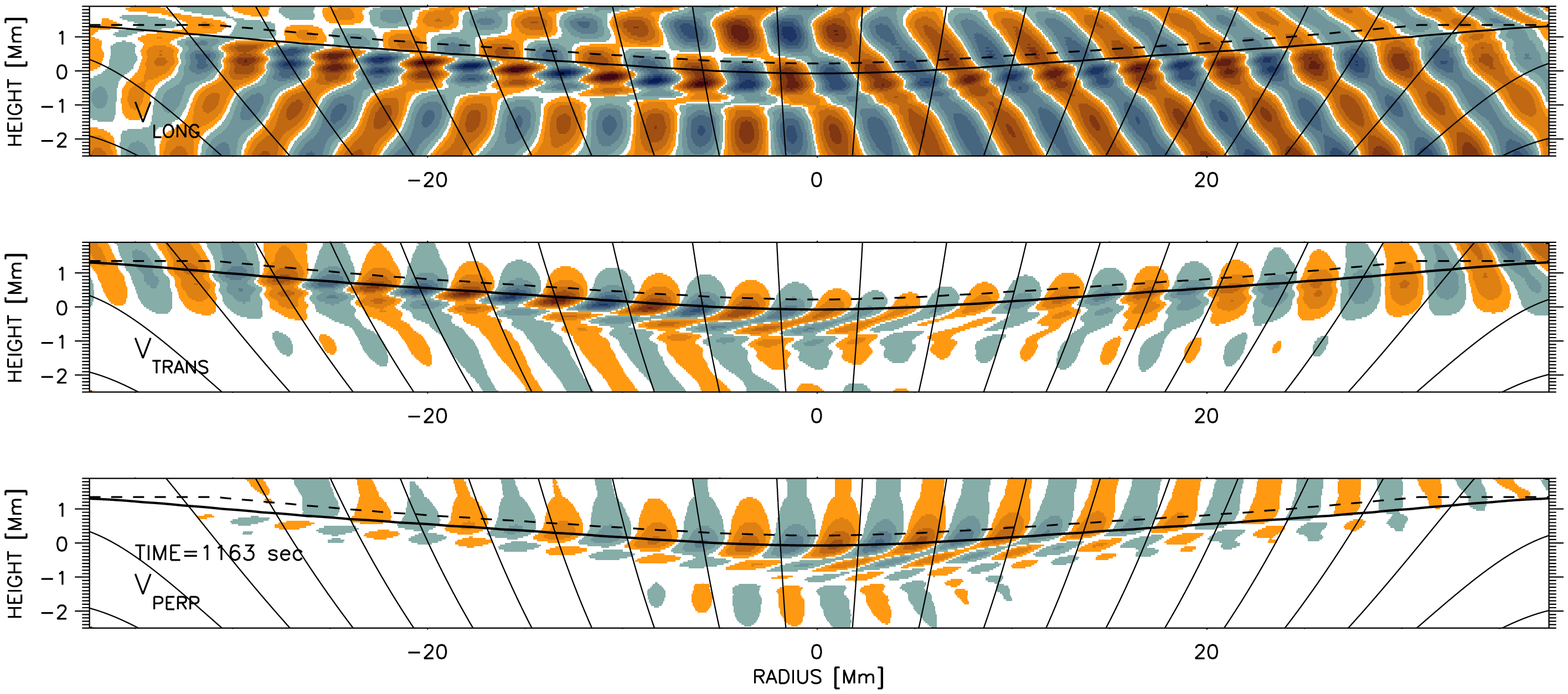}
\caption{Snapshots of the three orthogonal components of the
velocity taken at about 19 min after the start of the simulation
at $Y=7.5$ Mm from the sunspot axis. Upper panel: $V_{\rm long}$;
middle panel: $V_{\rm trans}$; bottom panel: $V_{\rm perp}$. The
blue-orange colors mean negative-positive velocity directions; the
range of the color coding is the same in the three panels. The
velocities are scaled with a factor of $\sqrt{\rho_0 c_S}$ (upper
panel) and $\sqrt{\rho_0 v_A}$ (two lower panels). Horizontal
solid line is the level where $c_S=v_A$; horizontal dashed line is
the fast mode reflection level. Magnetic field lines are inclined
black lines. The axes are not to scale. Note the presence of the
Alfv\'en mode in $V_{\rm perp}$ above $c_S=v_A$, where the fast
mode ($V_{\rm trans}$) is already reflected. Since the velocity
polarizations `long', `trans', and `perp' are based on $v_A\gg
c_S$ asymptotics, care must be taken not to over-interpret these
figures well below the equipartition level. The apparent
``discontinuity'' seen in $V_{\rm perp}$ near $Z=-1$ Mm (seen also
in Fig.~\ref{fig:vtime}) is a node. } \label{fig:vsnapshot}
\end{figure*}

To measure the efficiency of conversion to Alfv\'en waves near and
above the $c_S=v_A$ equipartition layer, we calculate the
time-averaged acoustic and magnetic energy fluxes as far as
possible from the conversion layer in the magnetically dominated
atmosphere ($v_A \gg c_S$):
\begin{equation}
\label{eq:fluxes}
\begin{split}
{\vec{F}_{\rm ac}} & =  \langle \delta P\,{\delta\vec{V}} \rangle \,,\\
{\vec{F}_{\rm mag}} & =   \langle \delta\vec{B} \times (
\delta\vec{V} \times \vec{B}_0 ) \rangle /\mu_0\,.
\end{split}
\end{equation}
The positive sign of the fluxes means energy propagating upwards.

We also calculate a measure of the time-averaged energy contained in all three
wave modes according to
\begin{equation}
\label{eq:energy}
\begin{split}
E_{\rm long} & =  \rho_0 c_S \langle \delta V_{\rm long}^2 \rangle \\
E_{\rm perp} & =  \rho_0 v_A \langle \delta V_{\rm perp}^2 \rangle \\
E_{\rm trans} & =  \rho_0 v_A \langle \delta V_{\rm trans}^2
\rangle\,,
\end{split}
\end{equation}
where in each case we use the corresponding velocity projections
into characteristic directions (Eq.~\ref{eq:directions}). For pure
acoustic and Alfv\'en waves, where there is strict equipartition
between kinetic and compressional or kinetic and magnetic energies
respectively, these would indeed be the true energies. Be that as
it may, these forms are convenient for purposes of exposition and
shall be used here.

\section{Velocity projections}\label{sec:vel}

Figure \ref{fig:vtime} shows an example of the projected velocity
components in our calculations as a function of height and time.
The velocities are taken at $X=10.5$ Mm, where the distinction
between the three wave modes in the magnetically dominated
atmosphere is clearly visible. In this representation the larger
inclination of the ridges means lower propagation speeds and vice
versa. Note, that projecting the velocities according to
Eq.~(\ref{eq:directions}) allows us to separate the wave modes only
in the magnetically dominated atmosphere, above the horizontal
solid line in Fig.~\ref{fig:vtime}.
The figure shows how the incident fast mode wave propagates to the
equipartition layer $c_S=v_A$ gradually changing its speed. After
reaching the equipartition layer at about 5 min into the
simulation, it splits into several components. The essentially
magnetic low-$\beta$ fast mode is produced above $Z=0.2$ Mm
(middle panel). This mode is reflected back down a few minutes
after it has been produced. The reflection height, calculated as
the height where the wave frequency $\omega$ and wave horizontal
wave number $k_x$ are related by $\omega=v_Ak_x$ (ignoring the
sound speed contribution to the fast wave speed), is well
reproduced in the simulations, as the velocity variations
associated with the fast mode decay rapidly above its reflection
layer.

\begin{figure*}
\center
\includegraphics[width=18cm]{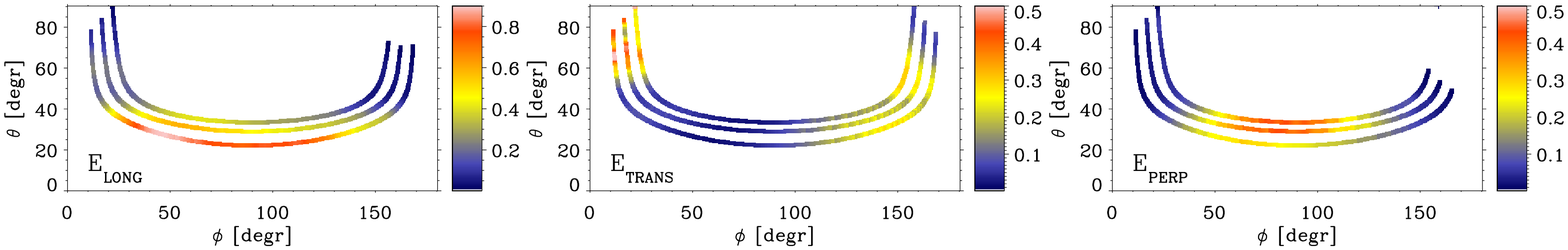}
\caption{Wave energies calculated according to Eq.~\ref{eq:energy}
at the top of the atmosphere (averages at heights from 0.4 Mm
above the $c_S=v_A$ layer up to $z=1.9$ Mm) for the three
projected velocity components in turn, selecting respectively
acoustic, fast, and Alfv\'en waves. The units of the color coding
are $10^6$ erg cm$^{-2}$ s$^{-1}$. The three strips in each panel
are the results of the three simulation runs at $Y=7.5$ (lower),
11.25 (middle) and 15 Mm (upper) from the sunspot axis.
% (from the outer to the inner).
}\label{fig:amplitudes}
%\vspace{0.5 cm}
\end{figure*}

Unlike the purely 2D case, there is an Alfv\'en mode produced
above the $c_S=v_A$ level (right panel). This mode has a clearly
distinct propagation speed compared to the acoustic mode (left
panel). The ridges of the Alfv\'en mode are nearly vertical since
the Alfv\'en speed at these heights is large (above 100 $\rm km\,s^{-1}$).

The low-$\beta$ essentially acoustic slow mode escapes to the
upper atmosphere tunnelling through the cut-off layer due to the
field inclination of $\theta$ = 30\degree\ that effectively
reduces the cut-off frequency by the factor $\cos\theta$.

The velocity amplitudes in Fig.~\ref{fig:vtime} are scaled with a
factor of $\sqrt{\rho_0 v_{\rm ph}}$, where $v_{\rm ph}=c_S$ for
the $V_{\rm long}$ component and $v_{\rm ph}=v_A$ for the other
two components. Although $v_A \gg c_S$ above the solid line in
Fig.~\ref{fig:vtime}, the scaled amplitudes of the Alfv\'en mode
are still smaller than of the slow acoustic mode at the selected
$X$ location.

Figure \ref{fig:vtime} also shows that the simulation enters into
a stationary stage after about 10-15 min. A snapshot of the
wave field developed in the stationary stage of the simulations is
given in Figure \ref{fig:vsnapshot}. The movie of this simulation
is available electronically.

The behaviour of the three wave modes changes across the sunspot
radial direction, as expected due to the change of the atmospheric
properties. The magnetic field is more inclined at the periphery
of the sunspot (much more so than is suggested by the figures,
since the vertical scale has been stretched relative to the
horizontal scale) and the azimuth varies from $\phi=0$\degree\ to
180\degree\ from the right to the left hand side of the sunspot.
The cut-off frequency reaches its maximum of 5.02 mHz at $X=0$,
$Z=0$ Mm, decreasing  steeply  above this due to the rise of
temperature in the chromosphere (Fig.~\ref{fig:sunspot}).

This steep temperature increase produces a partial reflection of
the slow acoustic mode in the magnetically dominated atmosphere.
This reflection is visible as an interference pattern formed above
the solid line in the $V_{\rm long}$ projection in the left part of
the sunspot (upper panel). This additional reflection also
produces stronger slow magnetic modes below the $c_S=v_A$ line in the
left part of the sunspot due to the secondary mode transformation
from the downgoing waves.

The fast magnetic mode reflection above $c_s=v_A$ is complete in
the central part of the sunspot, though the evanescent tail (the
exponentially decaying part of the fast wave above the classical
reflection point) still does not fully fit in the computational
box even here. At the periphery, the height of the simulation
domain is insufficient for the fast mode to complete its
reflection and it is partly absorbed by the PML boundary
condition. This produces artificially large fast mode amplitudes
at the periphery of the sunspot.

The Alfv\'en mode component generated after the transformation is
stronger at the horizontal locations where the fast wave is
reflected at lower heights ($X = 0$ -- $20$ Mm). The reason for a
this is apparent from Fig.~10b of \cite{CallyHansen2011}: the
fast-to-Alfv\'en conversion region is distributed throughout the
evanescent tail of the fast wave beyond its reflection point. At
the frequencies and wavenumbers typical of our simulations here,
that amounts to 20 scale heights or more! Since our computational
box is truncated only a few scale heights above $z_{\rm ref}$, we
can sample only a small part of the conversion process. Where
$z_{\rm ref}$ is lower (in the center of Fig.~3c) the Alfv\'en
conversion is of course enhanced relative to where it is higher
simply because our box includes more of the conversion region
there. Our experiments at 25 mHz, with $k_x$ also increased by a
factor of five, confirm this explanation, as for those waves the
evanescent tail is much shorter and the conversion region
therefore more compact. As expected, the higher frequency
simulations yield substantially higher Alfv\'en fluxes (see
Section \ref{sect:fluxes}).

\section{Energy of three wave modes}

We proceed by measuring the energies contained in each of the
projected velocity components at the upper part of our simulation
domain, above the Alfv\'en-acoustic equipartition layer $c_S=v_A$.
As the properties of the atmosphere change in the horizontal
direction across the sunspot, the height of the $c_S=v_A$ layer
changes as well (see Fig.~\ref{fig:sunspot}). To be consistent, we
take time averaged energies at heights from 400 km above the
$c_S=v_A$ layer up to the upper boundary of our simulation box in
the stationary stage of the simulations.

Figure \ref{fig:amplitudes} illustrates the energies as a function
of inclination and azimuth of the sunspot magnetic field lines at
the corresponding horizontal locations. The orientation of the
magnetic field is taken at heights where $c_S=v_A$ in each of the
simulations at three $Y$ locations. The three strips in each panel
of Fig.~\ref{fig:amplitudes} are for the three simulation runs.
The format of the figure allows comparison with the previous
studies, \cite{Cally+Goossens2008} Figure 2 and
\cite{Khomenko+Cally2011} Figure 4.

This figure demonstrates that the maximum energy of slow acoustic
waves is transmitted for inclinations around 30 degrees (left
panel). The dependence on the inclination is stronger than the
dependence on the azimuth. This result is consistent with previous
2D theoretical models of fast-to-slow mode transformation in the
homogeneous inclined magnetic field \citep{Crouch+Cally2003,
Cally2006, Schunker+Cally2006}. It is also consistent with the 3D
analysis by \citet{Cally+Goossens2008}. The explanation of this
effect is offered by the ray perspective. The fast-mode
high-$\beta$ waves launched from the sub-photospheric layers (with
an angle about 90 degrees, i.e. their lower turning point) reach
the Alfv\'en-acoustic equipartition layer with an angle close to
$20-30$ degrees \citep{Schunker+Cally2006}. Thus, for the
moderately inclined magnetic field the attack angle between the
wavevector and the magnetic field lines is small and the
transformation is efficient. Note also that the frequency of waves
is just at the cut-off frequency of the sunspot atmosphere, so due
to tunneling effects the slow mode acoustic energy is also present
over the wide range of inclinations.

\begin{figure*}
\center
\includegraphics[width=18cm]{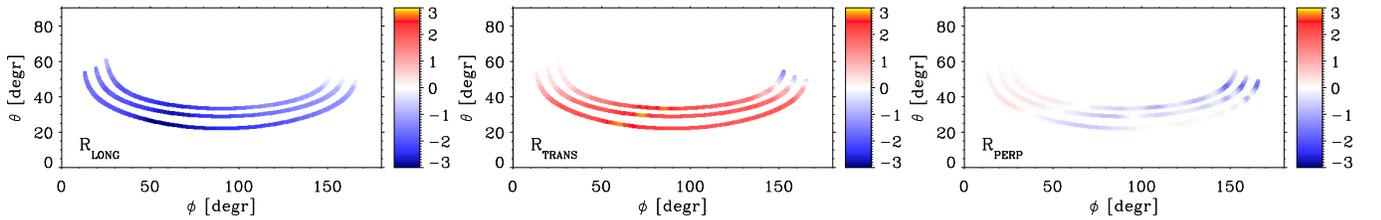}
\caption{Log$_{10}$ of the amplitude ratio $\delta
B/\sqrt{\mu_0\rho_0}$ to $\delta V$, both projected into the
characteristic directions according to Eq.~\ref{eq:directions}.
Left panel: slow acoustic mode ($\hat{e}_{\rm long}$); middle
panel: fast magnetic mode ($\hat{e}_{\rm tran}$); right panel:
Alfv\'en mode ($\hat{e}_{\rm perp}$). The three strips in each
panel are the results of the three simulation runs at $Y=7.5$
(lower), 11.25 (middle) and 15 Mm (upper) from the sunspot axis.}
\label{fig:ratio}
\end{figure*}

\begin{figure*}
\center
\includegraphics[width=12cm]{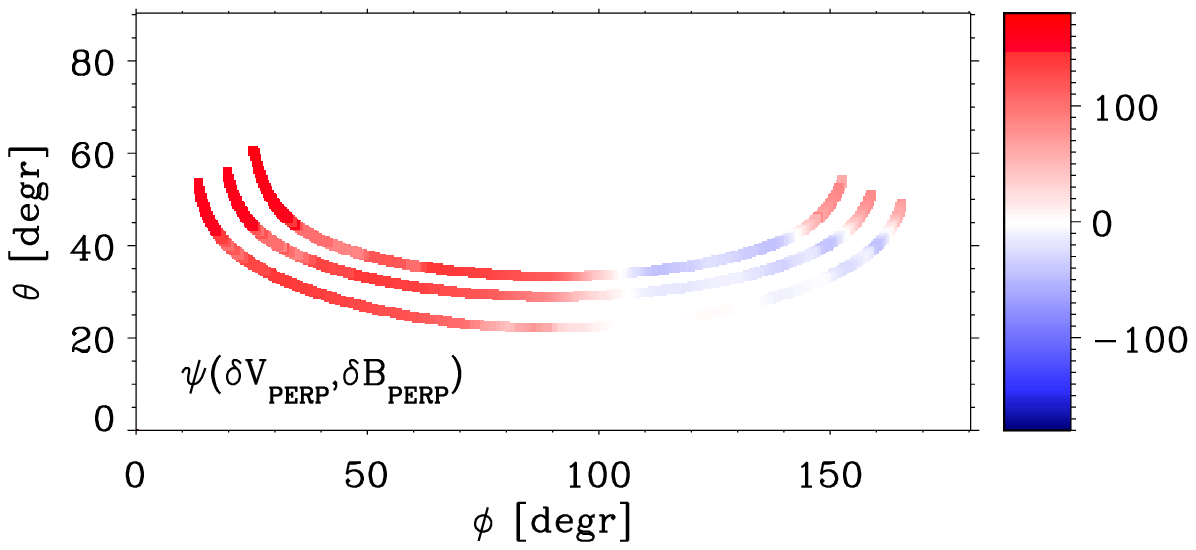}
\includegraphics[width=12cm]{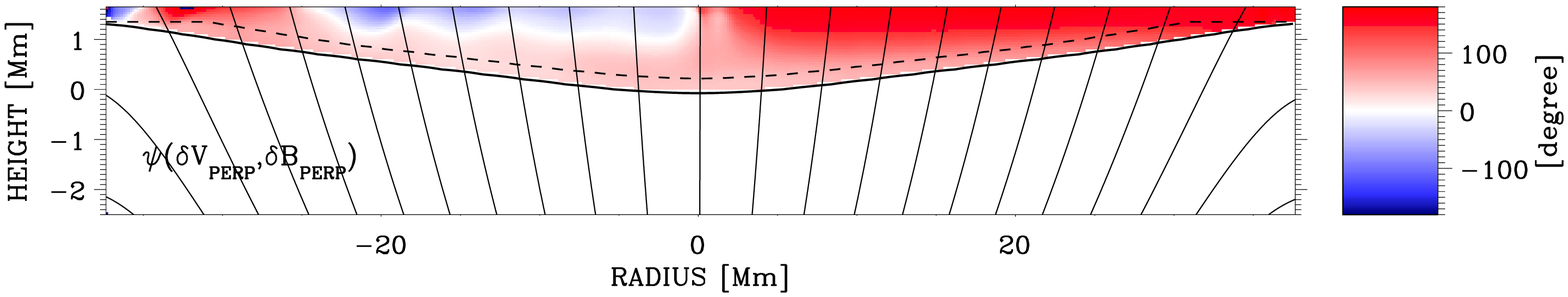}
\caption{Upper panel: phase shift between the variations of
$\delta V_{\rm perp}$ and $\delta B_{\rm perp}$ projections. The
format is the same as Fig.~\ref{fig:ratio}. Lower panel: same
quantity, but as a function of height and horizontal distance in
the simulation at $Y=7.5$ Mm from the axis. The phase shift is
only shown above the Alfv\'en-acoustic equipartition layer, where
the projections allow the separation between the modes and where
the Alfv\'en mode exists after the transformation.
}\label{fig:phase} \vspace{0.5cm}
\end{figure*}

The maximum energy of the Alfv\'en wave is present at inclinations
about 35\degree\ and azimuth between 70\degree\ and 100\degree\
(right panel of Fig.~\ref{fig:amplitudes}). The maximum power
increases from the outer to the inner strip indicating a tendency
towards larger transmission at larger inclinations. Thus, the
behaviour of waves in sunspot models is similar to the one found
previously in the models with homogeneous field
\citep{Cally+Goossens2008, Khomenko+Cally2011, CallyHansen2011}.
The transmission of Alfv\'en waves is more efficient for inclined
fields, which is important in the sunspot penumbra.

Some transmitted energy is also present in the $E_{\rm trans}$
component (middle panel of Fig.~\ref{fig:amplitudes}) near the
sunspot periphery, though we believe this to be an artefact of our
computational box not being tall enough to allow complete fast
wave reflection there.
%However, this energy corresponds to the fast magnetic waves in the
%upper atmosphere that could not complete their reflection at the
%periphery of the sunspot model, due to its insufficient height
%range. Thus, we believe that this power is rather a numerical
%artefact.

\section{Alfv\'en mode polarization relations}

To check if projecting the velocities in the directions given by
Eq.~(\ref{eq:directions}) gives an efficient way to separate the
Alfv\'en mode from the fast and slow magneto-acoustic modes, we
make use of the polarization relations for the Alfv\'en wave.
In a classical Alfv\'en wave variations of the magnetic field and
velocity should be related as \citep[see][]{Priest1984}:
\begin{equation}
\delta \vec{V}=\mp \frac{\delta \vec{B}}{\sqrt{\mu_0\rho_0}}\,,
\end{equation}
where the upper sign is for Alfv\'en waves propagating parallel to
the magnetic field and the lower sign is for those propagating in
the opposite direction. The kinetic and magnetic energy for an
Alfv\'en wave should be in equipartition, so the ratio $R=\delta
B/\sqrt{\mu_0\rho_0}/\delta V$ should be equal to one.

\begin{figure*}
\center
\includegraphics[width=14cm]{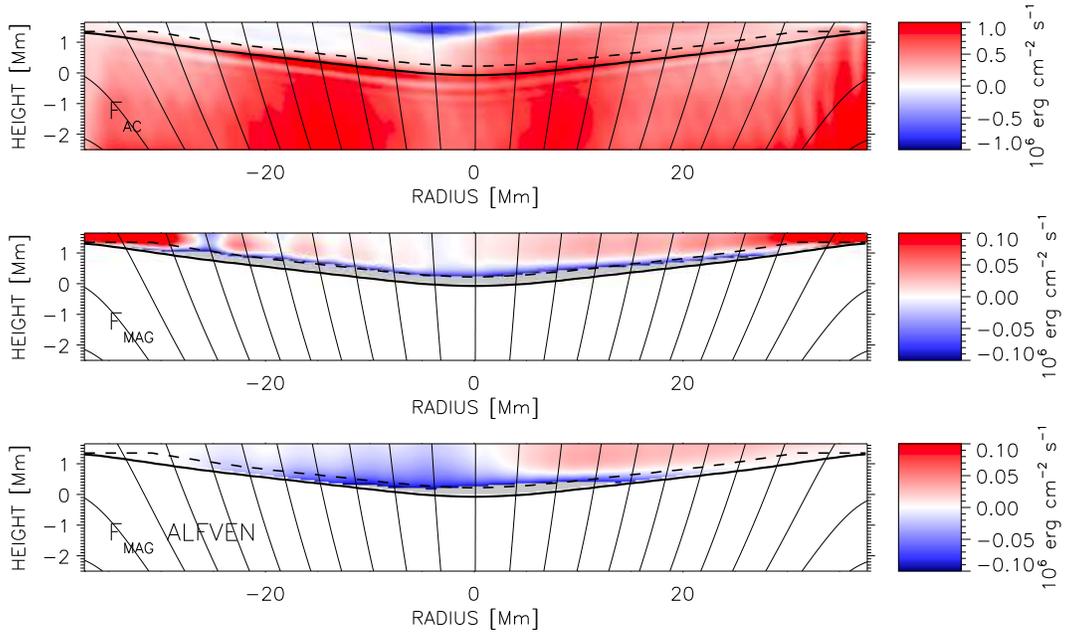}
\caption{Vertical component of the energy fluxes calculated after
Eq.~\ref{eq:fluxes}. Upper panel: acoustic flux; Middle panel:
magnetic flux; Bottom panel: magnetic flux, but using $\delta
B_{\rm perp}$, $\delta V_{\rm perp}$ in Eq.~\ref{eq:fluxes}, i.e.
magnetic flux due to Alv\'en waves. In the two bottom panel only
values above the Alfv\'en-acoustic equipartition layer are shown
where they are meaningful. Red colors mean upward flux, blue
colors mean downward flux. }\label{fig:fluxes}
\end{figure*}

To confirm the Alfv\'en nature of the transformed waves, we
checked the amplitude and phase relations for all three modes
reaching the upper atmosphere.
Figure \ref{fig:ratio} presents calculations of the amplitude
ratio $R$ for the three modes. In each of the cases $\delta B$ and
$\delta V$ pairs are the projections in the corresponding
characteristic direction for each mode (Eq.~\ref{eq:directions}).
The ratio is taken at the same heights as the energies from
Fig.~\ref{fig:amplitudes} and is averaged in time in the
stationary stage of the simulations.
The ratio turns out to be different for the three modes. In the
case of  the slow acoustic mode (left panel), the magnetic field
variations associated with the velocity variations are very small
(ratio $R$ about $10^{-2}$). In contrast, in the case of the fast
magnetic mode, magnetic field amplitudes are relatively strong
providing the ratio about $10^2$ (middle panel). For the Alfv\'en
mode the ratio stays around $10^0$ (right panel) meaning that the
velocity and magnetic field oscillations are in equipartition as
expected for an Alfv\'en wave.

Figure \ref{fig:phase} shows the phase shift between the $\delta
B_{\rm perp}$ and $\delta V_{\rm perp}$ corresponding to the
Alfv\'en mode. The situation here is very different from what one
might na{\"\i}vely expect. It is also different to the case of the
simple atmosphere and homogeneous magnetic field considered by us
previously in \citet{Khomenko+Cally2011}.
In the right part of the sunspot, for the azimuth between
0\degree\ and 90\degree\ the Alfv\'en waves mostly behave as they
should. The phase shift is about 180\degree, meaning an upward
wave propagation.
Nevertheless, in the left part of the sunspot ($\phi$ between
90\degree\ and 180\degree) the value of the phase shift indicates
a downward propagation. Taking into account that the amplitude
ratio (Fig.~\ref{fig:ratio}) gives clear evidence for the
Alfv\'enic nature of the waves, we conclude that for $\psi$
between 90\degree\ and 180\degree\ downward-propagating Alfv\'en
waves are generated by the mode transformation.

This accords perfectly with the uniform field modeling results of
\cite{CallyHansen2011} (see their Figures 4 and 5) indicating that
between {0\degree} and {90\degree} the upward propagating fast
waves couple most efficiently to upward Alfv\'en waves. Since an
alignment is needed between the direction of the wave propagation
and magnetic field for efficient coupling, where the azimuth is
between {90\degree} and {180\degree} the strongest coupling
happens between the downward propagating fast waves (after their
reflection in the magnetically dominated atmosphere) and the
Alfv\'en waves. In this case downward propagating Alfv\'en waves
are produced. We believe the same happens in the simulations,
producing downward propagating Alfv\'en waves in the left part of
the sunspot for $\psi$ above 90\degree.

All in all these calculations confirm the Alfv\'en nature of waves
selected by the $\hat{e}_{\rm perp}$ projection in our
simulations, including returning the predicted result regarding the
coupling between refracted fast waves and downward propagating Alfv\'en
waves in regions where the magnetic alignment is favourable.

\begin{figure*}
\begin{center}
\includegraphics[width=14cm]{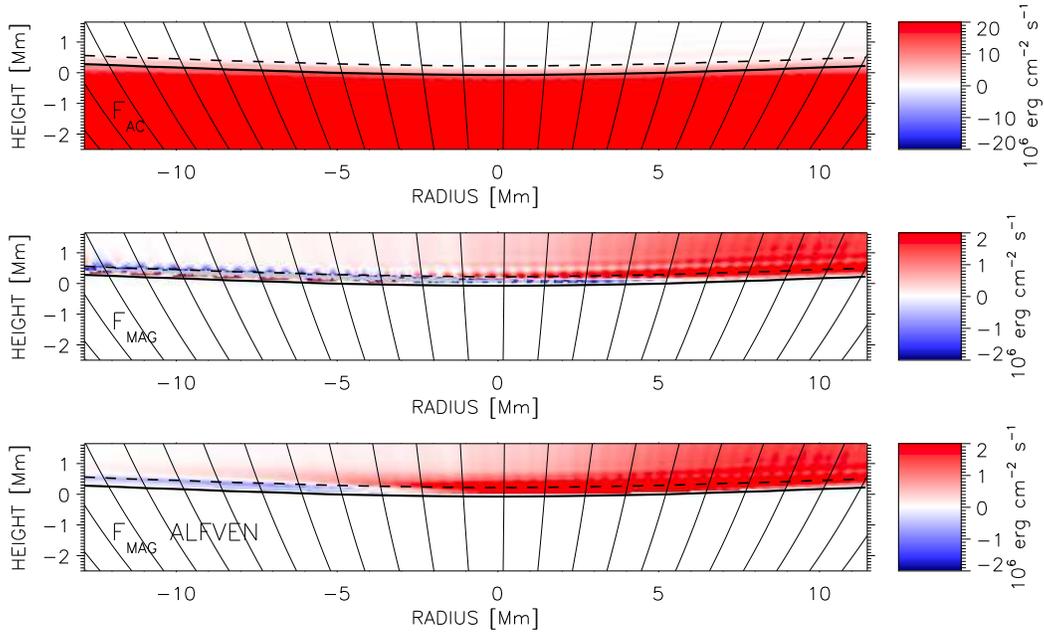}
\end{center}
\caption{Similar to Fig.~\ref{fig:fluxes}, but for a 25 mHz wave
with $k_x=6.85$ Mm$^{-1}$. The higher wavenumber shortens the
evanescent tail of the fast wave and therefore the Alfv\'en
conversion region is more compact and fits within the
computational box. A narrower box and three times finer horizontal
resolution has been used here to numerically resolve short-period
waves. } \label{fig:hifreq}
\end{figure*}

\section{Energy fluxes}
\label{sect:fluxes}

We proceed by calculating the acoustic and magnetic fluxes
according to Eq.~(\ref{eq:fluxes}). Figure~\ref{fig:fluxes} gives
the time averages of the vertical magnetic and acoustic fluxes
over the upper part of the simulation domain. In the case of the
magnetic fluxes we only show them above the {Alfv\'en-acoustic
equipartition} layer where they are produced and where the
separation between the modes according to
Eq.~(\ref{eq:directions}) is meaningful.

Above the equipartition layer, the upward acoustic flux is mostly
present in the right part of the sunspot for intermediate field
inclinations (upper panel). This result is in agreement with the
energy calculations from Fig.~\ref{fig:amplitudes} where the
maximum of the transmitted acoustic energy is for inclinations
about 30\degree.
In the left part of the sunspot the acoustic flux is about zero or
slightly negative. A large negative flux is also present at the
central part around the axis at the very top of the simulation
domain.
Thus, slow acoustic waves are reflected in the left part of the
sunspot. This appears to be in agreement with the visual impression from
Fig.~\ref{fig:vsnapshot} (upper panel) where the interference
pattern of the slow acoustic modes is observed to the left of
the axis. We believe that this partial reflection is due
to steep temperature increase around the axis in the upper
chromospheric part of the simulation domain. We have performed
additional simulations placing the upper boundary of the domain at
lower heights, below the chromospheric temperature increase. In
this case we did not observe downgoing slow acoustic waves. This
confirms that the reflection is a physical effect and not a
numerical artefact from the upper boundary condition of the
simulations.

The magnetic flux above the Alfv\'en-acoustic equipartition layer
(middle panel of Fig.~\ref{fig:fluxes}) is mostly positive. Note
that the absolute value of the magnetic flux is at all locations
lower than the acoustic flux. The exception is the peripheral
regions of the sunspot model where relatively large magnetic flux
is obtained. However, this artificially large flux is due to
insufficient height of the simulation box above the equipartition
layer making it impossible to complete the reflection for the fast
magnetic modes.

The magnetic flux in the middle panel of Fig.~\ref{fig:fluxes} is
due to a mixture of the fast and Alfv\'en modes. To separate the
flux of the Alfv\'en mode we calculated the magnetic flux from
Eq.~(\ref{eq:fluxes}) using the projections $\delta B_{\rm perp}$
and $\delta V_{\rm perp}$. The result is given in the bottom panel
of Fig.~\ref{fig:fluxes}.
In agreement with previous considerations we observe that there is
no significant flux in the peripheral part of the sunspot,
confirming its origin due to incomplete reflection of the fast
magnetic mode. As was already clear from the calculation of the
phase relations (Fig.~\ref{fig:phase}) the Alfv\'en flux in the
right part of the sunspot ($\psi$ between {0\degree} and 90\degree)
is positive and the flux in the left part of the sunspot ($\psi$
between 90\degree\ and 180\degree) is negative, indicating
downward propagating Alfv\'en waves. These downward propagating
waves are generated due to the coupling between the downgoing fast
magnetic mode (after its reflection) and the Alfv\'en mode.

The Alfv\'en flux only represents a minor fraction of the acoustic
flux in Fig.~\ref{fig:fluxes}. However, as explained at the end of
Section \ref{sec:vel}, we believe this to be an artefact of the
truncation of the broad Alfv\'en conversion region. To overcome
this limitation, we performed an experiment with five times larger
frequency ($\nu=25$ mHz) and horizontal wavenumber ($k_x=6.85$
Mm$^{-1}$) allowing us to significantly compact the conversion
region. This high frequency experiment is illustrated in
Fig.~\ref{fig:hifreq}, where indeed the upward Alfv\'enic flux in
the right half of the region now greatly exceeds the acoustic
flux, and there is significant but very compact downward Alfv\'en
flux in the left half.
Thus, the conclusion is that Alfv\'en flux can actually exceed
acoustic flux. Besides, Alfv\'en waves are more able to propagate
to great heights than are acoustic or fast waves, which are
limited respectively by shock formation and by reflection. This
suggests that they do indeed represent a significant product of
wave propagation and conversion in sunspots.

\section{Conclusions}

In this paper we have investigated the efficiency of conversion
from fast to Alfv\'en waves by means of 2.5D numerical simulations
in a complex sunspot-like magnetic field configuration.
It is important to realize that quantitatively simulating mode
transformation numerically is a challenge, as any numerical
inaccuracies are amplified in such second-order quantities as wave
energy fluxes. The tests presented in this paper prove the
robustness of our numerical procedure and offer an effective way
to separate the Alfv\'en from magneto-acoustic modes in numerical
simulations.

In general, the conclusions from the previous models in simplified
uniform magnetic field configurations \citep{Cally+Goossens2008,
Khomenko+Cally2011,CallyHansen2011} apply also to our more complex
sunspot-like magnetic field configuration, though there are some
apparent differences. We note three points in particular.

Firstly, the maximum of the magnetic energy of Alfv\'en waves
transmitted to the upper atmosphere is shifted toward more
inclined fields compared to the homogeneous field case.

Secondly, the amount of magnetic energy due to Alfv\'en waves
transmitted to the upper atmosphere is about 10 times lower than
for the acoustic waves. This differs from the conclusions reached
by \citet{Cally+Goossens2008} who find that at some particular
inclinations and azimuth angles the magnetic flux is larger. This
discrepancy is entirely to be expected, and is an artefact of the
limitations of our present simulations. As seen in Figure 10b of
\cite{CallyHansen2011}, the fast-to-Alfv\'en conversion region for
frequencies comparable to our 5 mHz is spread more or less
uniformly over some 20 scale heights from $z_{\rm ref}$ upwards.
Clearly, we do not have the luxury of such abundant space in our
simulation box, and so only a small fraction of the total
potential Alfv\'en flux is produced. In theory, this could be
remedied in any of several ways:

\begin{enumerate}
\item The box could be made substantially taller. However,
this would involve encompassing yet larger Alfv\'en speeds,
especially if a transition region is included, with unfortunate
consequences for our numerical time-step and therefore for the
practical feasibility of the calculation.

\item The magnetic field could be increased in magnitude. This is
an attractive course of action in any case, as the 900 G at $Z=0$
adopted here on the axis is very conservative, being a factor of 3
less than might be expected in a mature large sunspot. This would
lower both the $c_S=v_A$ equipartition surface and the $\omega=v_A
k_x$ fast reflection level  opening up more space above for mode
conversion. Of course, the maximum Alfv\'en speed would again
increase, with implications for the numerical time step.

\item Higher horizontal wavenumber waves could be modelled,
increased by say a factor of 5. As seen in \cite{CallyHansen2011},
Figure 10b, conversion of waves with $\kappa=k_x H\gtrsim1$ (where
$H$ is the density scale height) occurs in a considerably more
compact region near $z_{\rm ref}$. If frequency is increased by
the same factor, then $\omega/k_x$ and therefore $z_{\rm ref}$ is
unchanged, and the Alfv\'en conversion region should fit within
our computational box. Of course, little power is expected at such
high frequencies in the Sun, but nevertheless it is a useful
experiment to prove the point. Results of simulations on a similar
model but with $k_x=6.85$ $\rm Mm^{-1}$ and a frequency of 25 mHz
are shown in Fig.~\ref{fig:hifreq}, and are in accord with
expectations, yielding far higher Alfv\'en fluxes.

\end{enumerate}

Finally, where the reflecting fast waves meet similarly inclined
magnetic field in the conversion region as they propagate upward,
they preferentially transfer their energy to upgoing Alfv\'en
waves. However, if their upward path crosses field lines at large
angles there is little transfer. But then this correspondence may
be achieved on their way back down, in which case downward
propagating Alfv\'en waves are the beneficiaries of their largess.
This more efficient coupling to downward waves happens for azimuth
angles above 90\degree, and was predicted by
\citet{CallyHansen2011} based on their cold plasma ($\beta=0$)
analysis.

In our future studies we will extend this analysis to fully 3D
simulations and include both magnetic field twist and a
chromosphere-corona transition region. Of course twist contributes
to the field orientation in the conversion regions, and therefore
to the strength of mode conversion, both fast-to-slow and
fast-to-Alfv\'en, but we do not anticipate significant novel
effects beyond this. On the other hand, a transition region (TR)
is expected to be highly reflective to Alfv\'en waves, and
therefore to greatly reduce coronal Alfv\'en fluxes from
chromospheric mode conversion.  Results in a zero-$\beta$ model
\citep{HanCal11aa} confirm this, though they indicate that
substantial coronal Alfv\'en fluxes are still possible if the fast
wave reflection point is within a few chromospheric scale heights
of the TR, or indeed if the fast wave fails to reflect before it
reaches the TR. It will be interesting to test this in realistic
sunspot models.

Ultimately, our simulations will inform and be judged against
observations of MHD waves in solar magnetic structures. The
coronal observations of \citet{De-McICar07aa} and
\citet{TomMcIKei07aa} are useful indicators of power reaching that
height, though they are at the limits of our technology and
subject to line-of-sight integration ambiguities. Wave
observations at photospheric and chromospheric levels may be more
reliable and precise. In particular, correlations between velocity
and magnetic perturbation phases are crucial in disentangling the
various MHD modes in the low atmosphere \citep{norton01,
Khomenko+etal2003, SetSigMug02, FujTsun09}, though this is a very
difficult task even in full 3D simulations.

Significant progress has recently been made in the identification
of wave modes observed in sunspots and other solar magnetic
features. On the one hand, simultaneous photospheric and
chromospheric spectropolarimetric observations \citep[such as
those done with the TIP instrument on Tenerife, Canary Islands;
][]{Collados+etal2007} introduce the possibility of ``following''
wave propagation with height \citep{Centeno+etal2009,
Felipe+etal2010b} and obtaining phase shifts between oscillations
of magnetic field and velocity \citep{BellotRubio+etal2000}. On
the other hand, bidimensional spectrometers \citep[as e.g. IBIS
instrument;][or DOT telescope on La Palma, Canary
Islands]{Cavallini2000}  allow us to obtain high-resolution
2-dimensional velocity fields at different heights. Such
simultaneous observations are crucial for our understanding of the
physics of wave propagation in active regions. In sunspot umbrae,
where the magnetic field is predominantly vertical, slow
field-aligned magneto-acoustic waves are firmly detected and even
reproduced in simulations including their particular observed wave
pattern \citep{Centeno+etal2006, Felipe+etal2011}. In regions with
an inclined magnetic field (sunspot penumbra and network
canopies), a mixture of upward and downward propagating waves is
often discovered \citep{Braun+Lindsey2000, Veccio+etal2007,
Kontogiannis2010}. At photospheric level, these up- and downgoing
waves are, possibly, fast magneto-acoustic waves undergoing a
reflection process. In fact, from measured phase shifts between
velocity and magnetic field oscillations,
\citet{Khomenko+etal2003} found that the contribution of slow mode
waves is larger in umbral regions and the contribution of fast
mode waves becomes progressively more important toward the
umbra-penumbra boundary. Alfv\'en waves are usually detected
higher up in the corona \citep{De-McICar07aa, TomMcIKei07aa}. The
simulations presented here suggest that the complete decoupling
between the fast and Alfv\'en waves happens in the upper
chromosphere, or even above. Since the upcoming acoustic waves on
the Sun are uncorrelated, their attack angle with respect to the
sunspot magnetic field can be arbitrary, so a mixture of up- and
downgoing Alfv\'en waves will be produced, mostly above regions
with inclined magnetic field. Time variations of the chromospheric
magnetic fields, together with velocity oscillations, are strongly
desirable for the firm detection of Alfv\'en waves. However,
measuring magnetic fields in the chromosphere may be challenging
\citep[see, e.g.,][]{Manso+Bueno2010} and actual measurements are
scarce. As of today, very few measurements exist in active and
quiet chromospheric regions \citep[e.g.][]{Socas-Navarro2005,
TrujilloBueno+etal2005, Socas-Navarro2007, Centeno+etal2010,
Stepan+TrujilloBueno2010}. But, as far as we are aware, no time
variations have been reported. Further instrumental efforts,
developments of chromospheric diagnostic techniques, as well as
improved modelling, will be needed to constrain the results of the
presented study.

%%%%%%%%%%%%%%%%
\acknowledgments Financial support by the Spanish Ministry of
Science through projects AYA2010-18029 and AYA2011-24808 is
gratefully acknowledged. Work on this project began during a visit
by EK to Monash funded by an internal Monash Bridging Grant.

%\aareferences

\end{document}